# Guiding-center solitons in rotating potentials

Yaroslav V. Kartashov,[1] Boris A. Malomed,[2] and Lluis Torner[1]

[1]*ICFO-Institut de Ciencies Fotoniques, and Universitat Politecnica de Catalunya, Mediterranean Technology Park, 08860 Castelldefels (Barcelona), Spain*

[2]*Department of Interdisciplinary Studies, School of Electrical Engineering, Faculty of Engineering, Tel Aviv University, Tel Aviv, 69978, Israel*

We demonstrate that rotating quasi-one-dimensional potentials, periodic or parabolic, support solitons in settings where they are otherwise impossible. Ground-state and vortex solitons are found in defocusing media, if the rotation frequency, $\alpha$, exceeds a critical value. The revolving periodic potentials exhibit the strongest stabilization capacity at a finite optimum value of their strength, while the rotating parabolic trap features a very sharp transition to stability with the increase of $\alpha$.

Optical lattices (OLs) provide a unique setting for the creation, control, and manipulation of solitons in Bose-Einstein condensates (BECs) [1]. OLs can stabilize solitons against collapse in two- and three-dimensional (2D and 3D) BECs with attraction [2], and quasi-1D and quasi-2D lattices, which do not depend on one coordinate, support stable 2D and 3D solitons, respectively [3]. Axisymmetric Bessel potentials can stabilize 2D and 3D solitons too [4], and several types of 2D solitons were predicted in axisymmetric potentials which are periodic in the radial direction [5]. In optics, spatial solitons in a radial photonic lattice were observed in photorefractive crystals [6], where complex soliton trains can be created too [7]. In repulsive BECs, OLs give rise to gap solitons [8] and vortices [2,9]. Gap solitons were created in a $^{87}$Rb condensate [1], and were shown to remain stable at finite temperatures [10].

A new tool for the formation and manipulation of solitons in OLs is offered by *rotating optical lattices* (ROLs). They can be created using a revolving mask through which laser beam illuminates the condensate [11]; similar effective potentials may be induced by a rotating vortex lattice in a multicomponent BEC [12]. ROLs may be viewed as an extension of the concept elaborated in optics of periodic media [13], such as dispersion management [13,14] and tandem structures [15], where new objects emerge not possible in homogeneous media – i.e., *guiding-center solitons* [14]. Recently, 2D solitons were predicted in attractive BEC loaded in a square-shaped ROL [16]. At small rotation frequencies $\alpha$, an ordinary soliton may be trapped by the lattice, rotating together with it, while at large $\alpha$ the ROL supports ring-shaped solitons. Here, we aim to demonstrate a new possibility: ground-state solitons and localized vortices in *repulsive* BECs can be supported by rotating *quasi-1D* potentials, in the form of a lattice or parabolic trap. Note that, while the formation of fundamental solitons and vortices in rotating 2D potentials has been comprehensively investigated previously (see Refs. [11,12,16,17] and references therein), here we address a fundamentally different setting, based on a rotating quasi-1D potential in repulsive (self-defocusing) media. Unlike vortex lattices, *localized* objects in repulsive media cannot be supported by quasi-1D potentials without rotation. Besides BECs, a similar setting may be in principle realized in *twisted* defocusing layered optical media.

The evolution of the repulsive BEC under study is described by the normalized Gross-Pitaevskii equation for wave function $q$,

$$iq_\xi = -(1/2)(q_{\eta\eta} + q_{\zeta\zeta}) + \sigma |q|^2 q - pR(\eta,\zeta,\xi)q. \qquad (1)$$

Here, the repulsive nonlinearity corresponds to $\sigma = +1$, coordinates $\eta,\zeta$ are normalized to the transverse thickness of the condensate, $a_0$, $\xi$ is time in units of $ma_0^2/\hbar$, $m$ is the atomic mass, and $p$ measures the lattice depth in units of $\hbar^2/ma_0^2$. The quasi-1D lattice with period $2\pi/\kappa$, rotating at frequency $\alpha$ around the origin, corresponds to $R(\eta,\zeta,\xi) = \cos[\kappa\eta\cos(\alpha\xi) + \kappa\zeta\sin(\alpha\xi)] \equiv \cos[\kappa r\cos(\theta - \alpha\xi)]$, with polar coordinates $r$ and $\theta$. Using the scaling invariance of Eq. (1), we set $\kappa = 2$, while $\alpha$ and $p$ will be control parameters. A dynamical invariant of Eq. (1) is the norm, $U = \int\int_{-\infty}^{\infty} |q(\eta,\zeta)|^2 d\eta d\zeta$.

If the rotation is fast, i.e., $2\pi\alpha >> \kappa^2$, the potential approaches the averaged limit, expressed in terms of Bessel function $J_0$:

$$\langle R(r,\theta,\xi)\rangle_{\alpha\to\infty} \equiv (2\pi)^{-1}\int_0^{2\pi} \cos[2r\cos(\theta - \alpha\xi)]d\theta = J_0(2r). \qquad (2)$$

While, as said above, static quasi-1D OL cannot give rise to solitons in repulsive BEC, radial Bessel potentials support stable solitons [4], which suggests looking for similar structures stabilized by the ROL. First, we consider ROLs rotating around a potential minimum, i.e., Eq. (1) with $p > 0$. Simulations were conducted with the initial configuration displayed in Fig. 1(a), which is an exact soliton in the average Bessel potential. In the nonrotating quasi-1D lattice, the input quickly spreads out [Fig. 1(b)], while the rotation at moderate values of $\alpha$ results in retardation of the decay [Fig. 1(c)]. Further increase of $\alpha$ results in formation of a stable soliton, whose shape is almost identical to the input [Fig. 1(d)]. Thus, setting the quasi-1D OL into rotation qualitatively changes its trapping capability, resulting in the formation of average (*guiding-center*) solitons. To confirm the localization of these states, in Fig. 2(a) we show $U_r$, defined as the fraction of the initial norm which stays trapped within a circle of radius $2\pi/\kappa$ at $\xi = 128$, as a function of the rotation frequency. A transition from spreading of the input at small values of $\alpha$ to stable solitons at larger $\alpha$, in a region of width $\Delta\alpha \sim 3$, is evident. The width of the transition region slightly varies with the norm of the input.

Contrary to expectations that deeper ROL would provide stronger confinement, we have found that the dependence of $U_r$ on $p$ is *nonmonotonous* [Fig. 2(b)]. While a weak lattice cannot trap the atoms, the lattice potential which is too strong (for given $\alpha$) breaks the effective axial symmetry imposed by the rotation, and thus leads to leakage of atoms from the core area. Hence, there exists an *optimum* lattice strength, $p_{\mathrm{opt}}$, which grows with $\alpha$, while the steepness of the drop of $U_r(p)$ at $p > p_{\mathrm{opt}}$ decreases. We define a critical rotation frequency, $\alpha_{\mathrm{cr}}$, at which $U_r$ attains the level of $0.9U$ at $\xi = 128$. Figure 2(c) demonstrates that $\alpha_{\mathrm{cr}}$ increases with $p$, for $p > p_{\min} \approx 6$. Note that $p_{\min}$ is close to $p_{\mathrm{opt}}$, which was defined above as the value of $p$ providing for the maximum of $U_r(p)$. At $p = p_{\min}$, the soliton can be supported by the OL rotating at the smallest frequency [$\alpha_{\mathrm{cr}} \approx 8.9$ in Fig. 2(c)]. At $p < p_{\min}$, $\alpha_{\mathrm{cr}}$ abruptly increases and, at $p < 4.9$, the loss exceeds 10% even for $\alpha \to \infty$ [see Fig. 2(a)].

The rotating quasi-1D OL can also support ring-shaped *vortex solitons*. Figure 3 shows the transformation of the output shape of the vortex with the increase of $\alpha$, if the input [Fig. 3(a)] was taken as a vortex soliton in the average Bessel potential. While the vortex spreads out in the nonrotating OL [Fig. 3(b)], the vortex core tends to keep its annular shape in a slowly revolving lattice [Fig. 3(c)], and the entire localized vortex becomes stable at sufficiently large $\alpha$, see Fig. 3(d). The transition to the survival of the vortex, similar to that for the ground-state soliton, is revealed by dependence $U_r(\alpha)$ in Fig. 4(a), while Fig. 4(b) shows that $U_r$ is a non-monotonous function of the lattice strength. A specific feature of vortex solitons is that they are typically prone to azimuthal instability [18]. Nevertheless, we have found that the instability is prevented if the input norm exceeds a certain threshold.

To investigate the stability of the solitons in the limit of $\alpha \to \infty$, i.e., in average Bessel potential (2), we constructed them as $q = w(r)\exp(im\theta + ib\xi)$, with chemical potential $-b$. The family of the ground-state $(m = 0)$ solitons is characterized, in Fig. 5(b), by $U(b)$ dependences, the breakup points in these curves signaling a transition to a multi-ring shape in solitons with a large norm. Thus, soliton in Fig. 5(a) already slightly penetrates into first lattice ring. With increase of $U$ the part of soliton norm concentrated in the first lattice ring gradually increases, and soliton starts to penetrate into second and subsequent rings, developing multi-ring shape for large $U$. The ground-state and vortex solitons exist in respective domains, $0 < b < b_{\text{co}}^m$, with the cutoffs, $b_{\text{co}}^{m=1} < b_{\text{co}}^{m=0}$, increasing monotonously with $p$. To explore their linear stability, perturbed solutions were taken as $q = [w(r) + u(r)\exp(in\theta + \delta\xi) + v^*(r)\exp(-in\theta + \delta^*\xi)]\exp(ib\xi + im\theta)$, where $n$ and $\delta$ are the azimuthal index and instability growth rate of perturbation eigenmodes $u(r), v(r)$. We have found that the ground-state solitons are stable in the entire existence domain, which is consistent with the stability of the solitons in the ROL at $\alpha > \alpha_{\text{cr}}$. In contrast, vortex solitons are unstable against azimuthal perturbations with $n = 1$ near the existence border in the $(b, p)$ plane [Fig. 5(c)], where the norm of the vortex soliton drops below a certain threshold. By means of direct simulations, we have checked that counterparts of stable and unstable vortices in the Bessel potential, which corresponds to $\alpha \to \infty$, are, respectively, stable and unstable in ROLs at finite $\alpha$.

In the model with $p < 0$ in Eq. (1), which corresponds to the rotation pivot set at a local *maximum* of the quasi-1D lattice potential, it has been found that a rapidly rotating quasi-1D OL supports almost axisymmetric *ring-shaped* ground-state solitons trapped in the first minimum of respective Bessel potential (2). Properties of such solitons are similar to those described above. As the potential with a maximum at the center favors ring-shaped patterns, it readily supports vortex solitons. Accordingly, for given norm $U$ and lattice strength $|p|$, vortices lose their stability at much lower rotation frequencies than with $p > 0$. For instance, at $U = 32.3$ and $|p| = 25$, the critical frequencies are $\alpha_{\text{cr}} \approx 12$ and $\alpha_{\text{cr}} \approx 19$, for $p < 0$ and $p > 0$, respectively.

We have found stable solitons and vortices too in the rotating quasi-1D potential in the attractive medium, i.e., with $\sigma = -1$ in Eq. (1). We do not display those results here, as they are less surprising – as mentioned above, the quasi-1D lattice can stabilize solitons (although not vortices) without any rotation [3].

We also studied the soliton dynamics in the most fundamental rotating quasi-1D potential, *viz.*, a parabolic trap, corresponding to $-pR = (\Omega^2/2)[\eta\cos(\alpha\xi) + \zeta\sin(\alpha\xi) - \Delta]^2$ in Eq. (1). Here, $\Delta$ is a possible offset of the rotation pivot from the bottom of the parabolic trap, and, as the strength may be fixed by scaling, we set $\Omega^2 = 5$. While this potential cannot support solitons in repulsive condensates without rotation, at $\alpha \to \infty$ it assumes the averaged shape, $\Omega^2 r^2/4$, which suggests the existence of axisymmetric ground-state and vortex solitons. Transition from the decay of the input soliton in a slowly rotating trap, in Fig. 6(a), to the formation of a stable ground-state soliton in a rapidly rotating one, Fig. 6(b), is evident. A specific feature of the parabolic potential is that radiation can only escape in the unconfined direction, therefore no spiraling radiation tails emerge in this case.

We have also verified the stabilization for vortex solitons in the rotating parabolic potential. Before reaching a stationary shape, the vortices feature, at moderate values of $\alpha$, periodic splitting into two distinct pieces, followed by their recombination into an almost radially symmetric state, at a frequency close to $\alpha/2$, see Fig. 6(c). A similar periodic regime was found for vortex solitons in isotropic parabolic potentials, but with *attractive* nonlinearity [18].

As seen in Fig. 6(d), a noteworthy peculiarity of the rotating parabolic potential is a very sharp transition from the decay to stability with the increase of $\alpha$, with the width of the transient interval two orders of magnitude smaller than in the ROL, cf. Fig. 2(a). Within domains marked by circles in Fig. 6(d), the radiating ground-state solitons feature, on a longer time scale, a drift instability, which leads to a quick decay. Such instability may be readily understood. In the reference frame rotating along with the parabolic potential, the effective potential energy of a particle with mass $M$, which represents the soliton, is $W_{\text{eff}} = (M/2)[\Omega^2(\eta - \Delta)^2 - \alpha^2(\eta^2 + \zeta^2)]$, with the second term being the centrifugal energy. The equilibrium position at the origin, predicted by $W_{\text{eff}}$, disappears at $\alpha = \Omega$ (for $\Delta = 0$), which is close to frequencies at which the escape occurs. However, the soliton restabilization at larger $\alpha$, also observed in Fig. 6(d), cannot be explained within the framework of the quasi-particle approximation. Lastly, if the rotation pivot is set at a distance from the bottom of the parabolic trap, $\Delta \neq 0$, stable solitons perform oscillatory motion, with amplitude $A$ scaling as $\Delta$ and decreasing with $\alpha$. The proportionality of $A$ to $\Delta$ can be readily explained by the quasi-particle approximation.

Summarizing, we have demonstrated that setting quasi-1D potentials, lattices or parabolic traps, in rotation drastically changes their confining capabilities, allowing them to support stable solitons and localized vortices in repulsive BECs. The stabilization occurs above a threshold value of the rotation frequency. The rotating lattices exhibit the strongest confining capacity at an optimum value of the lattice strength, while parabolic traps feature a very sharp transition to the stability with the increase of the rotation frequency.

# Figure captions

Figure 1 (color online).     Evolution of the input (a), taken as the fundamental soliton of the associate Bessel potential, with norm $U = 6.4$, in the quasi-1D lattice with $p = 10$, rotating at frequency $\alpha = 0$ (b), 4.5 (c), and 20 (d). Here and in Figs. 3 and 6, the contour plots display the spatial distribution of $|q(\eta,\zeta)|$.

Figure 2.     (a) Norm $U_r$ versus the rotation frequency for $p = 10$. (b) $U_r$ versus lattice strength $p$ for initial norm $U = 12.9$. (c) The critical rotation frequency versus $p$ for $U = 12.9$.

Figure 3 (color online).     The evolution of the input vortex ring with norm $U = 17.2$ (a) in the revolving lattice with strength $p = 25$ and rotation frequency $\alpha = 0$ (b), 10 (c), and 30 (d).

Figure 4.     Norm $U_r$ versus the rotation frequency for $p = 25$ (a), and versus lattice depth for $U = 32.3$ (b).

Figure 5.     (a) Profiles of ground-state ($m = 0$) and vortex ($m = 1$) solitons at $b = 1.45$ and $p = 10$. (b) The norm of the ground-state solitons in Bessel potential versus $b$. (c) Existence domains of the solitons. Vortex solitons are unstable in the shaded area.

Figure 6 (color online).     Evolution of the input, chosen as the fundamental (a,b) or vortex (c) soliton of the averaged parabolic potential, with norm $U = 10$, in the quasi-1D parabolic trap rotating at frequency $\alpha = 2.2$ (a), 10 (b), and 2.4 (c). Panel (c) displays the stage of the periodic splitting with the largest separation between splinters, while dashed circles indicate inner and outer boundaries of input vortex, taken at level $0.5 \max|q|$. The duration of the recombination cycle in this case is 5.6. (d) The norm concentrated within the circle of radius $\pi$ at $\xi = 50$, versus the rotation frequency, for ground-state solitons with $U = 20$ (1) and 10 (2), and vortices with $U = 20$ (3) and 10 (4). In all cases, $\Omega^2 = 5$ and $\Delta = 0$.

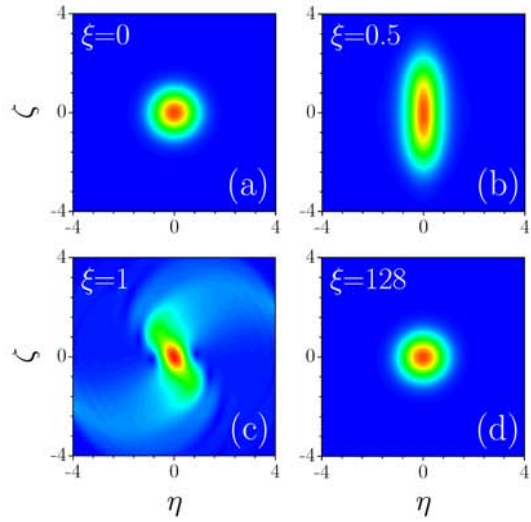

Figure 1 (color online). Evolution of the input (a), taken as fundamental soliton of the associate Bessel potential, with norm $U = 6.4$, in the quasi-1D lattice with $p = 10$, rotating at frequency $\alpha = 0$ (b), 4.5 (c), and 20 (d). Here and in Figs. 3 and 6, the contour plots display the spatial distribution of $|q(\eta,\zeta)|$.

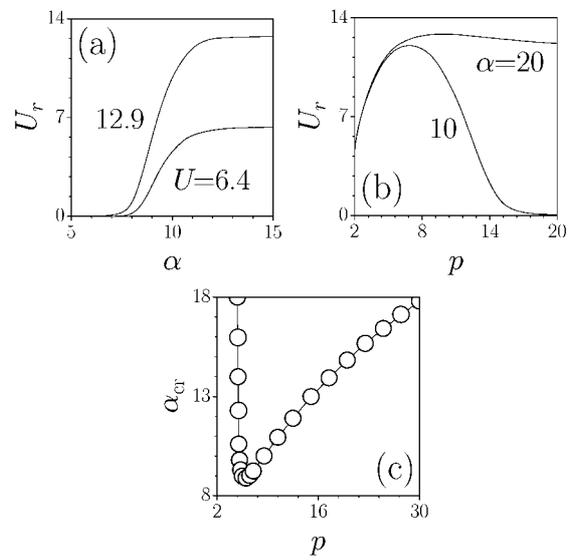

Figure 2. (a) Norm $U_r$ versus the rotation frequency for $p = 10$. (b) $U_r$ versus lattice strength $p$ for initial norm $U = 12.9$. (c) The critical rotation frequency versus $p$ for $U = 12.9$.

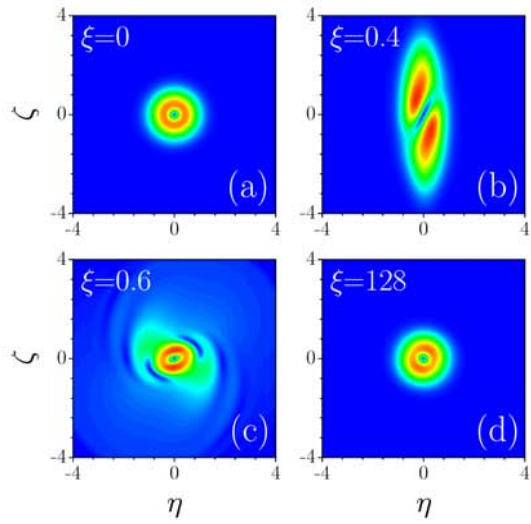

Figure 3 (color online). The evolution of the input vortex ring with norm $U = 17.2$ (a) in the revolving lattice with strength $p = 25$ and rotation frequency $\alpha = 0$ (b), 10 (c), and 30 (d).

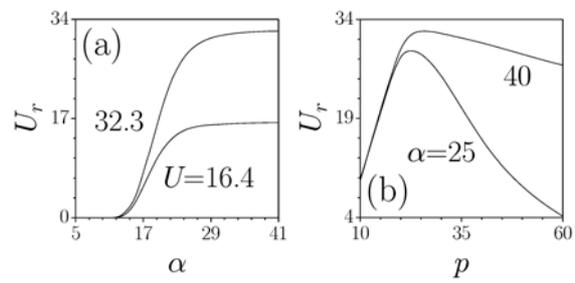

Figure 4.  Norm $U_r$ versus the rotation frequency for $p = 25$ (a), and lattice depth for $U = 32.3$ (b).

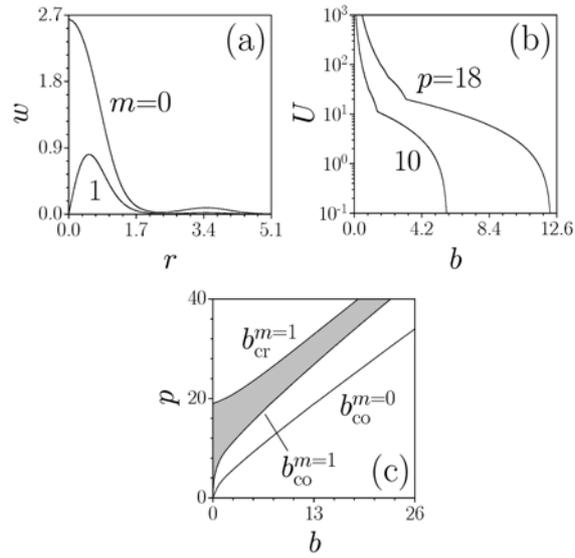

Figure 5. (a) Profiles of ground-state ($m = 0$) and vortex ($m = 1$) solitons at $b = 1.45$ and $p = 10$. (b) The norm of the ground-state solitons in Bessel potential versus $b$. (c) Existence domains of the solitons. Vortex solitons are unstable in the shaded area.

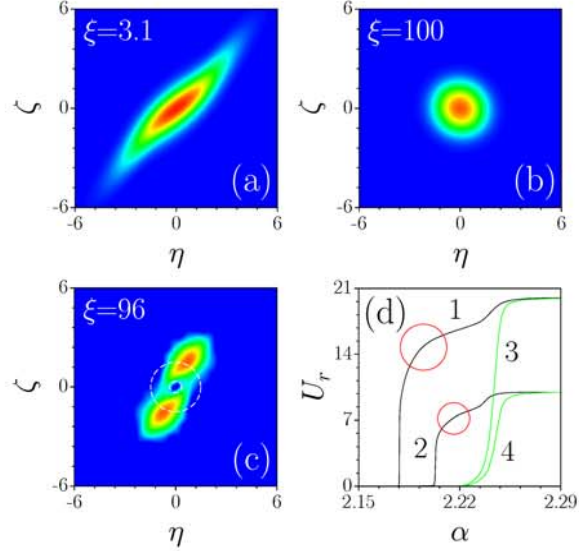

Figure 6 (color online). Evolution of the input, chosen as the fundamental (a,b) or vortex (c) soliton of the averaged parabolic potential, with norm $U=10$, in the quasi-1D parabolic trap rotating at frequency $\alpha = 2.2$ (a), 10 (b), and 2.4 (c). Panel (c) presents the stage of the periodic splitting with the largest separation between splinters, while dashed circles indicate inner and outer boundaries of input vortex (taken at level $0.5\max|q|$). The duration of recombination cycle in this case is 5.6. (d) The norm concentrated within the circle of radius $\pi$ at $\xi = 50$, versus the rotation frequency, for the ground-state solitons with $U=20$ (1) and 10 (2), and vortices with $U=20$ (3) and 10 (4). In all cases $\Omega^2 = 5$ and $\Delta = 0$.